# Ubiquitous Positioning: A Taxonomy for Location Determination on Mobile Navigation System

Wan Bejuri, Wan Mohd. Yaakob[1], Mohamad, Mohd. Murtadha[1] and Sapri, Maimunah[2]

[1]Faculty of Computer Science & Information Systems,
Universiti Teknologi Malaysia, Johor, Malaysia
wanmohdyaakob@gmail.com, murtadha@utm.my
[2]Faculty of Geoinformation & Real Estate,
Universiti Teknologi Malaysia, Johor, Malaysia
maimunahsapri@utm.my

## ABSTRACT

*The location determination in obstructed area can be very challenging especially if Global Positioning System are blocked. Users will find it difficult to navigate directly on-site in such condition, especially indoor car park lot or obstructed environment. Sometimes, it needs to combine with other sensors and positioning methods in order to determine the location with more intelligent, reliable and ubiquity. By using ubiquitous positioning in mobile navigation system, it is a promising ubiquitous location technique in a mobile phone since as it is a familiar personal electronic device for many people. However, there is an increasing need for better development of proposed ubiquitous positioning systems. System developers are also lacking of good frameworks for understanding different options during building ubiquitous positioning systems. This paper proposes taxonomy to address both of these problems. The proposed taxonomy has been constructed from a literature study of papers and articles on positioning estimation that can be used to determine location everywhere on mobile navigation system. For researchers the taxonomy can also be used as an aid for scoping out future research in the area of ubiquitous positioning.*

## 1. INTRODUCTION

Nowadays, the usage of GPS smart phone is increasingly widespread. It is because the capability of the smart phone can be used as personal navigator and communicator device. There are so many mobile navigation techniques which can be utilized to determine location which one of them is by using GPS which is already embedded in current GPS mobile phone. By using standalone GPS (ex: GPS smart phone), it is impossible to get better accuracy or signal especially in the particular obstructed environment (for ex: indoor car park, office, building, school and etc.). In addition, the object such as tree, high building, high wall and also people walking might be the contributors of the obstruction. These obstructions sometimes moved to another location which usually happened in indoor environment and finally make it difficult to estimate user's position. Moreover, the usage of other sensor on mobile phone such as WLAN, Bluetooth, GSM, and camera can be exploited to be alternative positioning sensor in order to determine user positioning in case if the GPS failed. Previous studies on mobile navigation system are focusing more on single sensor positioning and integration with external sensor. The integration with external sensor mostly is quite successful on positioning accuracy, but it is not really successful in terms of mobility. Moreover, the use of single positioning is also seems successful in mobility but however failed in ubiquity. The system's lacking of ubiquity is actually due to the lack of sensor integration within mobile phone. Thus, mobile positioning technologies need to be taxon or categorized together before the development of a reliable mobile navigation system.

The structure of the paper is as follows. Section 2 will present the reviews related work to taxonomy of location determination. Section 3 will present an overview of the proposed taxonomy. The detail of the proposed taxonomy will be covered in section 3.2. This will be followed by an exploratory of location determination taxon on radio frequency based, vision based and GPS based that are discussed in section 3.2, 3.2.1, 3.2.2, and 3.2.3. Finally, conclusions are given in section 4.

## 2. RELATED WORK

Most previous works focused on constructing taxonomies of location determination techniques by using specific type of positioning sensor. In an article describing the location systems for ubiquitous computing, Hightower *et al.,* (2001) [1] have developed a taxonomy for mobile computing devices in order to identify opportunities for new location-sensing techniques. Several evaluation properties have been listed: precision, accuracy, scale, cost, and limitations. However, taxonomy that was proposed by Hightower *et al.,* (2001) [1] has been criticised by Kjaergaard (2007) [2]. Kjaergaard (2007) [2] stated that it was not much help in specific question to radio location fingerprinting by proposing specific taxonomy for general properties of location fingerprinting systems which are: scale, output, measurements, and roles. Moreover, in an article describing the survey of wireless indoor positioning, Liu *et al.,* (2007) [3] have developed a taxonomy for performance of wireless indoor positioning based on [4] by listing: accuracy, precision, complexity, robustness, scalability and cost. Furthermore, the list was improvised by Gu et al. (2009) [5] in her article by introducing several evaluation criteria for assessing indoor positioning systems, namely security and privacy, cost, performance, robustness, complexity, user preferences, commercial availability, and limitations.

## 3. TAXONOMY

The proposed taxonomy is built around five (5) taxons listed each with definition in Table 1. This taxonomy is originated from Kjærgaard [2] that introduce taxonomy for radio location fingerprinting, but we enhance it by scoping for mobile positioning technologies (such as: GPS, Vision, WiFi, Bluetooth, and GSM). These were partly inspired by earlier work on taxonomies for location systems in general and from our literature study. The four taxons: scale, output, measurements, and roles describe general properties of mobile navigation systems.

Figure 1. Taxon Definition

| Taxon | Definition |
|---|---|
| Scale | Size of deployment area. |
| Output | Type of provided location information. |
| Measurements | Types of measured input signal from positioning sensor. |
| Roles | Division of positioning system architecture |
| Estimation Method | Algorithm that is used to estimate positioning information from measured input signal. |

The focus of the proposed taxonomy is on methods for location determination in mobile positioning technology that can be used on mobile navigation system. The evaluation properties such as in article Hightower [1] and Lie *et al,* [3] will be not covered in this paper.

## 3.1. General Taxon

The general taxons that proposed in Kjærgaard [2] are improved based on our scope. These taxons are shown in Figure 2 including subtaxons. In this following sections, taxons are presented up to four (4) references are given to papers or articles that propose systems that are grouped below the particular taxon.

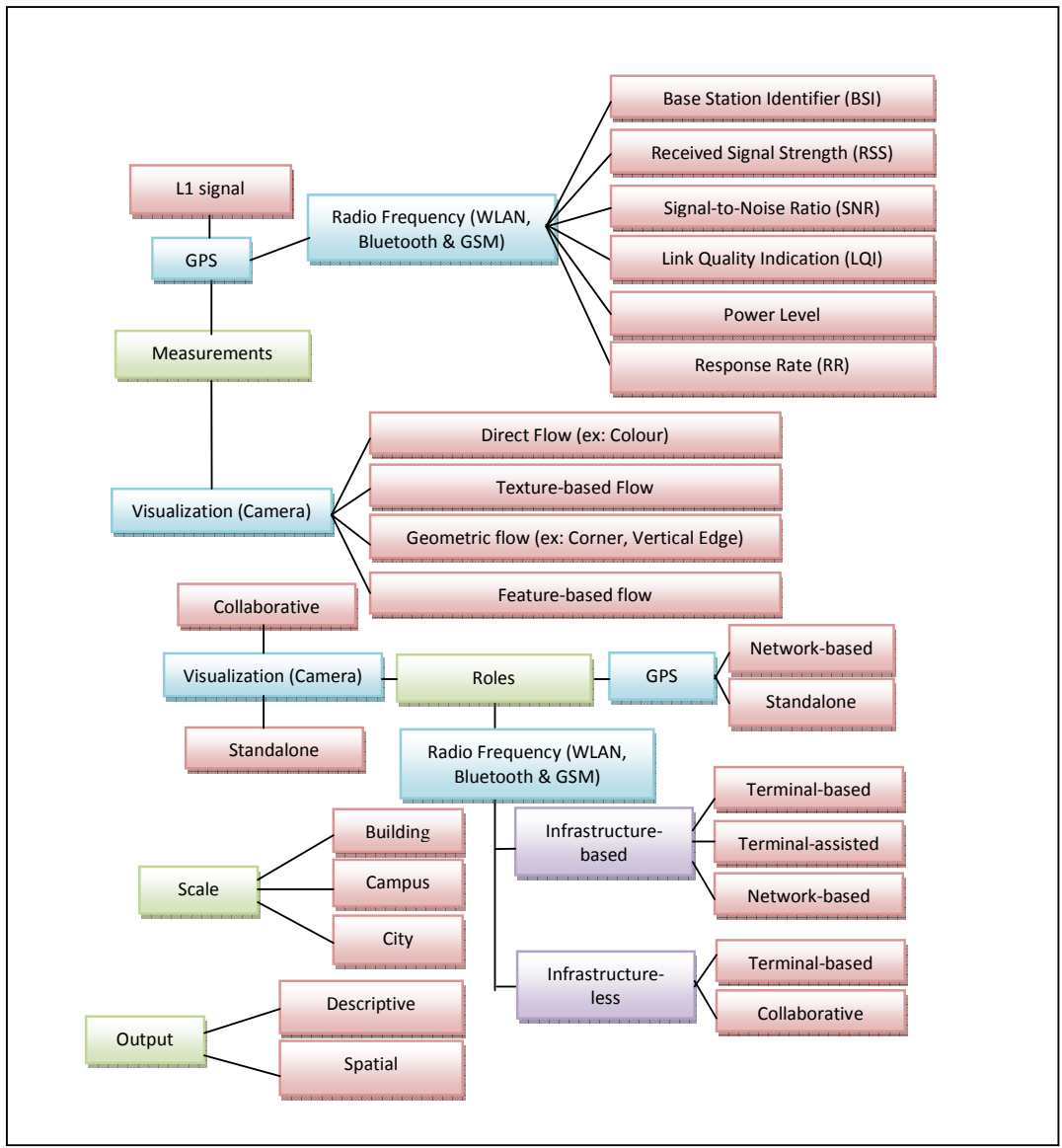

Figure 2. Scale, Output, Measurements and Roles (Kjærgaard [2], Sun *et al,* [6], Raviv *et al,* [7], Savarese *et al,* [8], Gavrila *et al,* [9], Rohrmus [10], Liapis *et al,* [11], Post *et al,* [12], Blake *et al,* [13], Aponte *et al,* [14], Neri *et al,* [15], Manodham *et al,* [16], Norouzi *et al,* [17], Aktas *et al,* [18])

Figure 2 displays the four (4) factors necessitating identification, which are scale, measurement, output and roles. Scale refers to the size of a system's deployment, which is essential as it affects data collection and the scale constraints of certain systems due to particular assumptions. Output refers to the categories of location information, of which there are two. The first is descriptive locations, which are expressed in the form of object-allocated variables or identifiers; the second is spatial locations, which are expressed as a set of coordinates that correspond with a spatial reference system.

Measurements refer to the type of measured input signal from positioning sensor. For radio frequency positioning (WLAN, Bluetooth and GSM), the measurement technique can be used such as: Base Station Identifiers (BSI), Received Signal Strength (RSS), Signal-to-Noise Ratio (SNR), Link Quality Indicator (LQI), Power Level (PL), and Response Rate (RR). BSI is a name distinctively allotted for a base station, while RSS, SNR, and LQI are radio-obtained signal propagation metrics utilized to manage and optimize communication. PL signifies the information regarding the present sending power that is sent by the signal sender, and RR refers to the frequency of obtained measurements sent by a particular base station throughout a certain temporal period. For GPS, L1 is needed in order to obtain input signal measurement. For vision technology,   direct flow, texture-based flow, geometric flow, and feature-based flow can be utilized. This vision measurement is a low level image detection technique.

Roles explain the allocation about division of positioning system architecture. As for radio frequency positioning such as WLAN, Bluetooth and GSM, it is presented in two (2) types; the first is infrastructure-based systems, which rely on a pre-installed powered infrastructure of base stations. The second is known as infrastructure-less system that encompass of ad-hoc-installed battery-powered wireless clients, with some undertake the function of base stations. For GPS, it is presented in two (2) types which are standalone and network-based. Standalone is a system architecture which involves single GPS observation, while network-based involves more than one (1) GPS observation. For vision technology, it is presented in two (2) types which are standalone and collaborative. Standalone is a system architecture which involves single camera observation, while collaborative involves more than one (1) camera observation. Usually collaborative technique used involves three (3) dimension measurement or for reducing error.

## 3.2. Location Estimation Taxon

Figure 2.3 depicts the location estimation method used for predicting locations. However, it is very challenging to taxonomize all possible methods because nearly all methods developed for machine learning (Sun *et al,* [6]) (see Witten *et al,* [19] for a list of methods) or in the field of estimation (see Crassidis *et al,* [20] for a list of methods) are applicable to the problem of location estimation. On another note, numerous researches have also been carried out regarding location estimation previously. In this paper, we follow Liu *et al,* [21] for radio frequency positioning, Bonin-Font *et al,* [22] for camera navigation and Quddus *et al,* [23] for GPS.

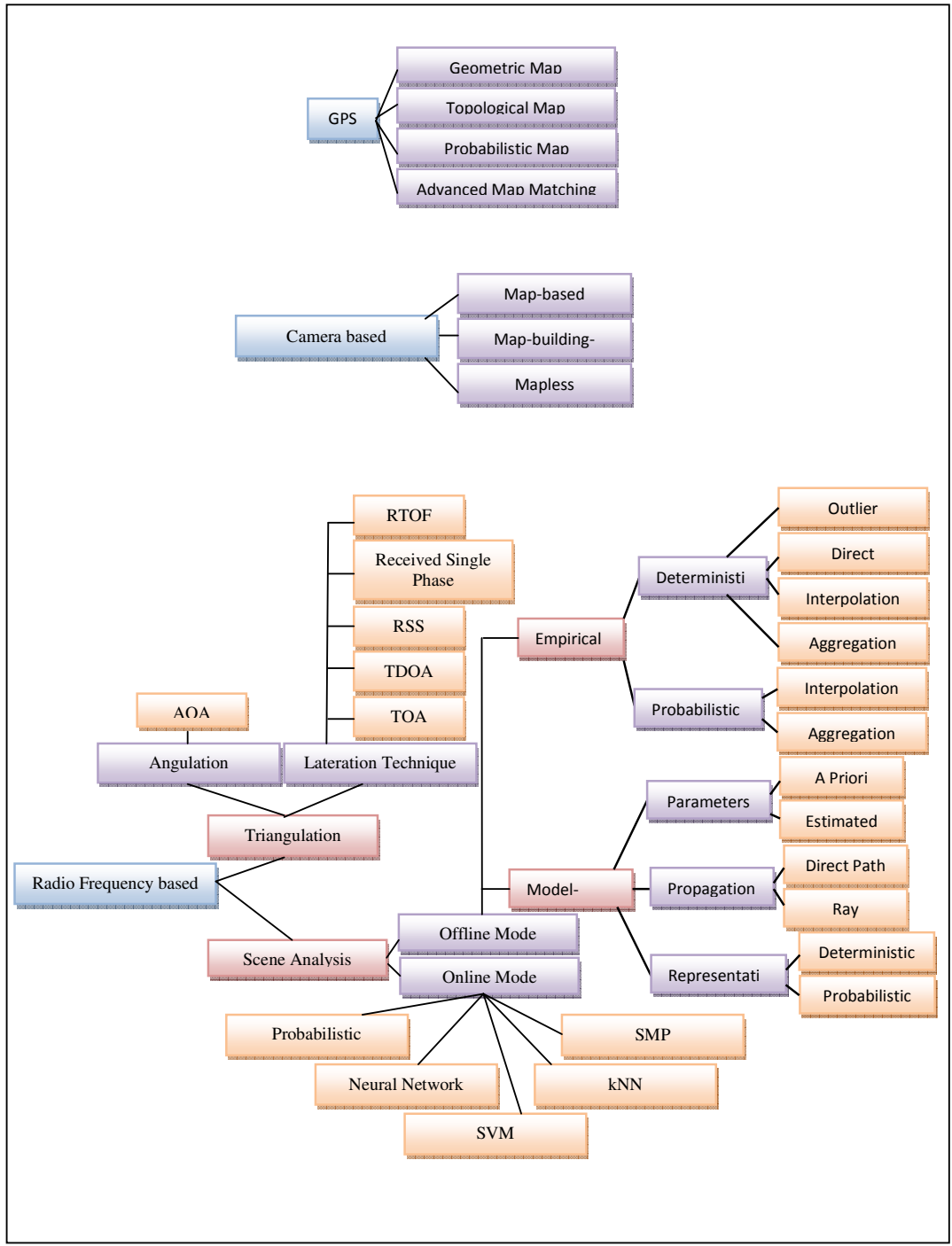

Figure 3. Location Estimation Method (Kjærgaard [2], Liu *et al,* [21], Bonin-Font *et al,* [22], Quddus *et al,* [23])

### 3.2.1. Radio Frequency Based

Radio frequency positioning estimation technique is split into the two (2) categories of scence analysis and triangulation. Scene analysis refers to the type of algorithms that first collect features (fingerprints) of a scene and then estimate the location of an object by matching online measurements with the closest *a priori* location fingerprints. There are two phases for location fingerprinting: offline stage and online stage (or run-time stage). A site survey is performed in an environment during the offline stage. The location coordinates/labels and respective signal strengths from nearby base stations/measuring units are collected (see algorithm example in figure 3). During the online stage, a location positioning technique uses the currently observed signal strengths and previously collected information to figure out an estimated location (see algorithm example in figure 3). Triangulation uses the geometric properties of triangles to estimate the target location. It has two derivations: lateration and angulation. Lateration estimates the position of an object by measuring its distances from multiple reference points (see example algorithm). So, it is also called range measurement techniques. Angulation locates an object by computing angles relative to multiple reference points (see example algorithm at Van Veen *et al,* [24], Stoica *et al,* [25], Ottersten *et al,* [26]).

### 3.2.2 Camera Navigation Based

Location estimation for camera navigation is split into the three (3) categories; map-based, map-building-based and mapless. Map-based consists of providing the database with a model of the environment in the system. These models may contain different detail degree, varying from a complete computer-aided design (CAD) model of the environment to a simple graph of interconnections or interrelationships between the elements in the environment. The technique can be considered as self-localization and is a fundamental technique for a correct navigation. The main steps in the technique are: capture image information, detect landmarks in current views (objects, edges or corner), match observed landmarks with those contained in the database according to certain criteria, and update the user position, as a function of the matched landmarks location in the stored map.

Map-building-based is method that provides system capability by exploring the environment and building its map by itself (see example in Erhard et al, [27] and Goedemé *et al,* [28]). The navigation process starts once the system has explored the environment and stored its representation.  Mapless is a method that provides the system in which navigation is achieved without any prior description of the environment. It is depends on the elements observed in the user environment such as: walls, features, doors and desks. There are two (2) main navigation techniques based on mapless: optical-flow and appearance-based navigation. Technique based on optical-flow is estimate the features or objects motion in the sequence of images. Most of the researchers develop an optical flow system by using pioneering techniques based on (see in article Hom *et al,* [29] and Lucas *et al,* [30]). While the technique that using appearance-based is a matching techniques based on the images storage in a previous recording phase. These images are then used as templates. The system self locates and navigates in the environment to match the current viewed frame with the stored templates. Outdoor navigation can be categorized in two (2); structured and unstructured environment. Algorithm color based that already developed by Rizzi *et al,* [31],  Stanikunas *et al,* [32], Ebner  [33], Foster [34], Ebner [35], Biancoa *et al,* [36] and Martínez-Verdú *et al,* [37], and texture based can be used in 'road following method' in order to recognize the lines of the road or any structured infrastructure and navigate consistently. For unstructured environment, method such as Wilcox *et al,* [38] and Krotkov *et al,* [39] can be applied since there are no regular properties to solve kind situation (see detail in Bonin-Font *et al,* [20]).

### 3.2.3   GPS Based

In this section, we focus only on location determination by using standalone embedded GPS on mobile phone in structured environment (ex: road) since it needs integration with other device or sensor in order to make it survive in unstructured environment. This technique called as map matching algorithm and it can be subdivided into geometric analysis, topological analysis, probabilistic map-matching algorithms and advanced map-matching algorithms.

A geometric map-matching algorithm makes use of the geometric information of the spatial road network data by considering only the shape of the links (see detail in Greenfeld [40]). In the geometric map-matching algorithm, the technique based on simple search algorithm is most commonly used. There are three (3) types, which are point-to-point matching, point-to-curve matching and curve-to-curve matching. Point-to-point matching refers to the matches of each position fixes to the closest 'node' or 'shape point' of a road segment. Point-to-curve matching (see detail in Bernstein *et al,* [41] and White *et al,* [42]) refers to the matches of the point (the position fix obtained from the navigation system) on to the closest curve in the network. Meanwhile, curve-to-curve matching (see detail in Bernstein *et al,* [41] and White *et al,* [42]) compares the vehicle's trajectory against known roads.

A topological map-matching algorithm is an algorithm that using the links geometry (such as: points, lines, and polygons) as the links connectivity and contiguity (see example in Greenfeld [40]).

A probabilistic algorithm is refers to algorithm that requires the information definition of an elliptical or rectangular confidence region around a position fix which is obtained from a navigation sensor (such as: GPS). The error region is superimposed on the road network to identify a road segment. If the error region contains more than one street, this algorithm will perform a weighted search on the candidate streets.

Advanced map-matching algorithms are referred to as those algorithms that use more refined concepts such as a Kalman Filter or an Extended Kalman Filter (e.g., Kim et al, [43]), Dempster-Shafer's mathematical theory of evidence 2 (e.g., Yang *et al,* [44]), a flexible state-space model and a particle filter (e.g.,  Gustafsson *et al,* [45]), an interacting multiple model (e.g., Cui *et al,* [46]), a fuzzy logic model (e.g., Kim *et al,* [47] ), or the application of Bayesian inference (e.g., Pyo *et al,* [48]).

### 4. CONCLUSIONS

In the next generation of mobile navigation system, people will need more alternative types of context information of the environments on the mobile phone, not just only limited to communication services.  One of the information context which is focused in this paper is the location context in the mobile phone. The usage of location context (ex: GPS) is becoming more popular nowadays. By utilizing and improving existing location determination techniques, this will enable location wares to be more intelligence in order to upgrade the quality of life. In this paper, we presented the taxonomy for location determination on mobile navigation system which is crucial to the establishment of ubiquitous positioning on mobile phone. The proposed taxonomy was constructed from a literature study of papers and articles about positioning estimation. The taxonomy was presented consists of the five taxons as follows: scale, output, measurements, roles and estimation method. Valuable taxonomies can account for everything that is known so far and can predict things to come, as variations of parameters accounted for and enumerated in the taxonomy. The proposed taxonomy was presented shows the depth and the breadth of our understanding. We would like others to join and based on the inputs from the community we can further improve the proposed taxonomy.

# ACKNOWLEDGEMENTS

This paper was inspired from my master research project which is related to ubiquitous positioning on mobile phone. The author also would like to thank Mohd Murtadha b Mohamad for his insightful comments on earlier drafts of this paper.

**Authors**

Wan Mohd Yaakob Wan Bejuri received the Dip. in Electronic Engineering in 2005 from Politeknik Kuching Sarawak and the B.Sc. (Computer Science) from Universiti Teknologi Malaysia in 2009. He currently M.Sc. candidate at Universiti Teknologi Malaysia.

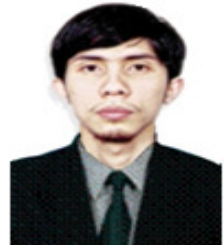

Mohd Murtadha Mohamad received the B.Eng. (Computer) from Universiti Teknologi Malaysia. He also received Ph.D. (Electrical) and M.Sc. (Embedded Sys. Eng) from Heriot-Watt University. He currently senior lecturer at Universiti Teknologi Malaysia and also a member of IEEE. For detail, please go to: http://csc.fsksm.utm.my/murtadha/

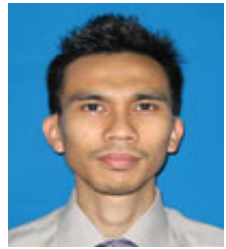

Maimunah Sapri  received the Dip. in Estate Management from Institute of Technology MARA. She also received B.Surv. (Hons.) Property Management and M.Sc. (Facilities Management) from Universiti Teknologi Malaysia and reveived Ph.D (Real Estate Management) from Heriot-Watt University. She currently senior lecturer at UniversitiTeknologi Malaysia.

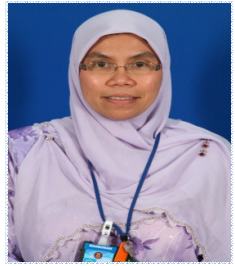